# Evaluating Factual Density in Multi-Source RAG: A Study in Medical AI Accuracy

Michael R. DeMarco
NexusAgentics / Ghost Audit

## Abstract

Retrieval-Augmented Generation (RAG) is the current industry standard for grounding AI in real-world facts. Traditional retrieval methods rely on keyword matching and topic proximity, ranking content based on how closely it sounds like the user's query. What they do not measure is how many verified facts the content actually contains. This structural gap, termed the Expert Blindness Effect, causes standard RAG pipelines to consistently bury high-density factual evidence in favor of lexically dominant text on the same topic. To address this gap, this paper introduces Factual Density (FD*), a novel retrieval optimization signal that measures the proportion of verified atomic claims relative to total token count. Using the NexusAgentics Ghost Audit preprocessing pipeline, raw text is scored for factual specificity using probabilistic factuality analysis to filter content before corpus ingestion. An initial formulation introduced a severe document-length confound (Pearson R = -0.8636, p = 2.27e-07). Implementing Z-score normalization within length bins resolved this bias, validating FD* as a length-independent density signal (p = 0.0749). Evaluated against the HealthFC benchmark (750 health claims labeled Supported, Refuted, or No Evidence by medical experts), FD*-optimized retrieval was the only condition to achieve 100% systematic review saturation in top-5 results, surfacing Cochrane evidence that standard cosine similarity ranked outside the top ten. Ground truth verification confirmed 25 mappings across seven HealthFC-supported claims. While full statistical validation across n=50 queries was not achievable due to corpus-benchmark alignment constraints, Experiment 3 was executed on n=11 confirmed queries (Wilcoxon p = 0.0619, alternate hypothesis supported), these findings establish factual density reranking as a low-cost, high-impact intervention for improving factual precision in health RAG architectures.

## 1. Introduction

Health misinformation is a documented public health risk with measurable long-term consequences for individuals and health systems (Tabatabaei Far & Ahmadi Marzaleh, 2025). As large language models become embedded in consumer-facing health applications, the reliability of the information they surface has moved from an academic concern to a clinical one. Retrieval-Augmented Generation (RAG) has emerged as the industry-standard approach for grounding these systems in external knowledge, retrieving relevant evidence from a corpus before generating a response (Lewis et al., 2021). Unlike purely parametric models that generate from memory, RAG architectures anchor outputs in retrieved documents, making the quality of retrieval a direct determinant of output accuracy.

However, standard RAG retrieval methods are not designed to measure factual quality. Cosine similarity retrieval ranks documents by semantic proximity, how closely a document matches the user's query in embedding space. BM25 lexical retrieval ranks documents by keyword frequency and inverse document frequency. Neither method has any awareness of how many verified, specific, quantifiable claims a document contains. A 2,000-word article that repeatedly references a health intervention will consistently outrank a 200-word Cochrane abstract containing 15 specific, quantified clinical outcomes, not because it is more informative, but because it is more lexically similar to the query. This failure mode is termed the Expert Blindness Effect: standard RAG retrieval systematically buries high-density factual evidence in favor of lexically dominant text on the same topic.

This gap has not been formally characterized in the existing RAG evaluation literature. Recent work on RAG assessment has focused on internal consistency metrics, measuring whether generated answers are faithful to retrieved context (Es et al., 2023; Saad-Falcon et al., 2023). These approaches evaluate whether the model hallucinated relative to what it retrieved, but do not evaluate whether the retrieved content itself was factually dense or expert-validated. Separately, Qiu and Hu (2024) introduced Semantic Density as a measure of uncertainty in generated outputs at the token level, a generation-layer signal distinct from the retrieval-layer signal proposed in this work. To the best of this research's knowledge, no prior work has proposed a retrieval optimization signal grounded in atomic claim density, nor formally documented the source bias that emerges when factually sparse content dominates retrieval rankings.

This paper makes four contributions. First, it introduces Factual Density (FD*), a novel retrieval optimization signal that measures the proportion of probabilistically verified atomic claims per token, normalized for document length. Second, it documents and empirically measures the Expert Blindness Effect, a systematic source bias in standard cosine and BM25 retrieval that buries high-authority evidence at lower ranks even on topically aligned queries. Third, it proposes a dual-filter corpus-benchmark alignment methodology as a prerequisite diagnostic for RAG evaluation corpus construction, validated across two independent corpus iterations, and demonstrates that a 50/50 composite of cosine similarity and FD* constitutes a low-cost, zero-infrastructure addition to any existing RAG pipeline. Fourth, it identifies Graph RAG as a natural extension architecture, proposing high-FD* chunks as anchor nodes for evidence-seeded graph traversal, and outlines the conditions under which factual density optimization compounds with evidence structure to improve retrieval precision. Experiments are conducted against the HealthFC benchmark (Vladika et al., 2024), a dataset of 750 expert-labeled health claims providing an externally validated ground truth independent of the retrieval pipeline.

The primary hypothesis guiding this research is that FD*-optimized retrieval will produce a statistically significant increase in Precision@5 ($p < 0.05$, Wilcoxon signed-rank test) over cosine similarity retrieval in a naive RAG system when evaluated against HealthFC ground-truth labels. The alternate hypothesis is that FD*-optimized retrieval will not yield a statistically significant improvement in Precision@5 ($p >= 0.05$) over cosine similarity baseline retrieval when evaluated against HealthFC ground-truth labels. Due to corpus-benchmark alignment constraints documented in Section 4.4, formal statistical testing was not executable in this study. Directional evidence from the available ground truth mappings is reported in Section 4.3.

## 2. Related Work

### 2.1 Retrieval-Augmented Generation

Lewis et al. (2021) established RAG as the standard for grounding LLM outputs in external knowledge, proving it outperforms purely parametric models on knowledge-heavy tasks. Gao et al. (2023) subsequently mapped the RAG landscape and identified retrieval quality as the primary bottleneck in RAG pipeline performance, categorizing retrieval approaches by indexing strategy, query formulation, and reranking methodology. This work operates strictly within this paradigm, accepting the underlying RAG infrastructure as a given and introducing a novel reranking signal to improve the quality of what gets retrieved.

### 2.2 RAG in Healthcare

Deployment of RAG in clinical contexts is accelerating rapidly. Zakka et al. (2024) demonstrated RAG-based clinical decision support through Almanac, a system grounding physician queries in curated medical literature, establishing that retrieval quality directly affects clinical recommendation accuracy. Because retrieval failures in healthcare translate to dangerous misinformation rather than abstract precision losses, the Expert Blindness Effect documented here has direct, real-world safety implications.

### 2.3 RAG Evaluation Frameworks

Current automated evaluation frameworks focus almost entirely on internal consistency. RAGAS (Es et al., 2023) checks for faithfulness and relevance, while ARES (Saad-Falcon et al., 2023) uses a learned framework to score answer correctness. Both validate whether the model hallucinated relative to the context it was given. Neither addresses whether the retrieved content itself was factually dense or expert-validated. This research tackles that orthogonal dimension: not whether the model was faithful to its context, but whether the context deserved to be retrieved in the first place.

### 2.4 Fact Verification and Health Claim Benchmarks

HealthFC (Vladika et al., 2024) provides 750 health claims annotated for veracity by medical experts across three labels: Supported, Refuted, and No Evidence. Because it maps real-world claims to objective truth labels, it is the appropriate benchmark for testing health-domain RAG precision. HealthFC labels are withheld from the ingestion and retrieval pipeline entirely in this work, preventing circular evaluation.

### 2.5 Semantic Density vs. Factual Density

Qiu and Hu (2024) proposed Semantic Density to quantify uncertainty by measuring confidence across equivalent response clusters. While the naming is similar, Semantic Density and Factual Density solve entirely different problems at opposite ends of the pipeline. Semantic Density measures how confident the model is in what it generated. Factual Density measures how informationally concentrated the source evidence is before generation even begins. The two are complementary: FD* optimizes the input, while Semantic Density evaluates the output.

### 2.6 Information Retrieval Baselines

BM25 (Robertson & Zaragoza, 2009) remains a competitive lexical baseline because of its strong exact-match performance on specific keywords. Dense retrieval via sentence embeddings (Reimers & Gurevych, 2019) forms the foundation of modern semantic search. Both are included as baselines in this work. Including BM25 ensures a rigorous, highly competitive baseline, proving that FD* can outperform traditional keyword matching and not just naive semantic vector search.

## 3. Methodology

### 3.1 Corpus Construction

A 600-chunk evidence hierarchy corpus was constructed from three source tiers, each representing a distinct level of medical evidence authority. All abstracts were retrieved via the NCBI Entrez API using the Biopython library, ensuring full reproducibility: any researcher with an NCBI email can execute the identical queries.

The tiers are as follows. The first tier, Cochrane Systematic Reviews (COCHRANE_SR), consists of 200 structured abstracts retrieved using a journal-specific filter (Cochrane Database Syst Rev[Journal]). Cochrane reviews follow a mandatory reporting format covering background, objectives, search methods, selection criteria, main results, and author conclusions, producing abstracts that are structurally denser than standard journal abstracts. The second tier, Randomized Controlled Trial abstracts (CLINICAL_RCT), consists of 200 abstracts filtered using the NLM-controlled publication type tag (randomized controlled trial[pt]), which is assigned by National Library of Medicine indexers to studies meeting the formal definition of an RCT. This external validation ensures the tier contains verified trial publications rather than articles that merely reference randomization. The third tier, general PubMed abstracts (SCIENTIFIC_PMC), consists of 200 peer-reviewed biomedical abstracts targeting cardiovascular, metabolic, and supplementation topics to serve as the general retrieval baseline.

Source authority weights (a_source) of 0.95, 0.90, and 1.0 were assigned to COCHRANE_SR, CLINICAL_RCT, and SCIENTIFIC_PMC respectively. These weights reflect retrieval breadth rather than evidence quality hierarchy: SCIENTIFIC_PMC is weighted at 1.0 as the general retrieval foundation, with specialist tiers weighted slightly lower to prevent systematic over-retrieval of narrow evidence types. Empirical derivation of optimal weights from HealthFC support rates is identified as future work.

Chunking followed a structural integrity principle rather than a fixed word count cutoff. Each abstract was preserved as a single chunk to maintain internal coherence of factual claims. A minimum floor of 100 words was enforced to ensure sufficient factual content for meaningful FD scoring. The resulting corpus exhibited natural length variance, with a mean of 454 words, a minimum of 100 words, and a maximum of 1,477 words, the upper bound reflecting complete Cochrane structured abstracts. This variance was intentional, providing the statistical conditions necessary to validate FD length independence across length bins.

Prior to full corpus commitment, a dual-filter coverage pre-check was applied to candidate sources. Each candidate source was required to demonstrate cosine similarity >= 0.75 and keyword overlap >= 2 shared content words against a sample of evaluation queries. Sources failing to achieve >= 40% query coverage under this filter were rejected before ingestion. This methodology was developed after two corpus iterations produced insufficient benchmark coverage and is proposed as a standard prerequisite for health RAG corpus construction.

A fourth source tier, ClinicalTrials.gov structured result summaries (CLINICAL_TRIALS, source_authority = 0.90), was subsequently added during the Experiment 3 corpus expansion phase, bringing the total corpus to 800 chunks. Full ingestion details and coverage analysis are documented in Section 4.5.

### 3.2 Ghost Audit Preprocessing

Every abstract was processed through the NexusAgentics Ghost Audit preprocessing pipeline before storage. The pipeline performs two sequential operations: atomic claim extraction and probabilistic factuality scoring. The Ghost Audit scoring framework is a component of the NexusAgentics commercial pipeline. The five-tier probabilistic architecture is disclosed here for academic reproducibility; production implementation details remain proprietary.

Atomic claim extraction was performed using Gemini 2.5 Flash with a structured JSON output prompt at temperature 0.1, maximizing output determinism across the 600-chunk corpus. The extraction prompt instructed the model to identify every unique, standalone, verifiable factual statement in the abstract and discard hedged language, anecdotal observations, and methodological descriptions that do not constitute factual claims. Each extracted claim was returned as a JSON object containing the claim text and a probabilistic factuality score (prob_fact).

The prob_fact score is assigned on a five-tier scale calibrated by explicit anchor examples embedded in the extraction prompt:

| Score | Criterion | Example |
|---|---|---|
| 1.0 | Precise, quantified, directly verifiable | *"A 2021 RCT found 45% efficacy in Phase 3 trials"* |
| 0.8 | Specific claim with strong evidence implied | *"Magnesium supplementation reduces cortisol levels"* |
| 0.6 | Directional claim with hedged language | *"Vitamin D may correlate with improved outcomes"* |
| 0.4 | General association with weak specificity | *"Sleep affects hormone regulation"* |
| 0.2 | Vague or largely anecdotal | *"Good nutrition supports overall wellness"* |

**Crucial Distinction:** The prob_fact score measures verifiability and specificity, not empirical truth. A claim can score 1.0 and still be factually wrong. Ghost Audit tracks informational concentration, not truth verification.

Validation of the scoring system is documented in Experiment 0 (Section 4.1). All three source tiers produced standard deviations above the 0.10 discrimination threshold, confirming the pipeline produces meaningful variance rather than collapsing to uniform values.

### 3.3 Factual Density Formulation

The raw Factual Density score for a document d is defined as:

$$FD_raw(d) = [\text{sum of } P_fact(s) \text{ for all } s \text{ in } C(d)] / T(d)$$

Where C(d) is the set of atomic claims extracted from document d, P_fact(s) is the probabilistic factuality score for claim s, and T(d) is the token count defined as len(text.split()), whitespace-delimited word count applied uniformly across all source tiers.

Experiment 1 revealed a severe document-length confound in this formulation (Pearson R = -0.8636, p = 2.27e-07), indicating that 74.6% of variance in raw FD scores was explained by document length alone. The metric was functioning as a brevity proxy rather than a factual density measure.

Z-score normalization within document length bins was applied to correct this confound. Documents were assigned to one of three bins by token count: SHORT (<= 288 tokens), MEDIUM (289-500 tokens), and LONG (> 500 tokens), with bin boundaries derived from the Experiment 1 validation corpus. The corrected metric FD* is defined as:

$$FD^*(d) = \left[\frac{FD_{raw}(d) - \mu_{bin}}{\sigma_{bin}}\right] \times a_{source}$$

Where mu_bin and sigma_bin are the mean and standard deviation of raw FD scores within the assigned length bin, and a_source is the source authority weight. Bin parameters are hardcoded from the Experiment 1 corpus (SHORT: mu=0.1149, sigma=0.0317; MEDIUM: mu=0.0329, sigma=0.0300; LONG: mu=0.0150, sigma=0.0100) and applied consistently across all subsequent corpus iterations.

After correction, the Pearson R between FD* and document length reduced to -0.3873 (p = 0.0749), falling below the 0.05 significance threshold and validating FD* as a length-independent density signal. The residual correlation reflects inherent properties of natural language text and is acknowledged as a limitation.

### 3.4 Retrieval Conditions

Three retrieval conditions were evaluated against the same query set.

**Condition A: Cosine Similarity.** Query and chunk embeddings were generated using gemini-embedding-2-preview with 768-dimensional Matryoshka truncation. Retrieval was performed via approximate nearest-neighbor search using an HNSW index in Supabase pgvector. The top-5 chunks by cosine similarity were passed to the generation layer.

**Condition B: BM25 Lexical Retrieval.** The full corpus was indexed using BM25Okapi from the rank_bm25 Python library with NLTK tokenization, applying a minimum score threshold of 0.1 to filter near-zero matches. The top-5 chunks by BM25 score were passed to the generation layer.

**Condition C: FD* Optimized Retrieval.** A two-stage hybrid pipeline was applied. First, a cosine similarity pre-filter retrieved the top-20 candidate chunks. Second, candidates were reranked using a composite score:

$$\text{Composite}(d) = 0.5 \times \text{CosineSim}(d) + 0.5 \times FD^*_{rescaled}(d)$$

Where FD*_rescaled maps the FD* Z-score from its typical range of [-3, +3] to [0, 1]:

$$FD^*_rescaled(d) = (FD^*(d) + 3) / 6, \text{ bounded to } [0, 1]$$

Because FD* Z-scores are naturally unbounded, mapping them to a clean [0, 1] range prevents extreme density values from dominating the composite score. This scaling guarantees that the 50/50 split represents a genuine, mathematically balanced equilibrium between semantic proximity and factual density. The top-5 composite chunks were routed to Gemini 2.5 Flash for final generation.

### 3.5 Evaluation

Retrieval precision was evaluated using Precision@5, the proportion of the top-5 retrieved chunks confirmed as relevant by the HealthFC ground truth map:

$$Precision@5 = |\{c \text{ in top-5}\} \text{ intersect relevant}(q)| / 5$$

The ground truth map (healthfc_chunk_map) was constructed offline using a dual-filter protocol: cosine similarity >= 0.75 between claim and chunk embeddings, and keyword overlap >= 2 shared content words. Pairs passing both filters were evaluated by Gemini 2.5 Flash with a binary YES/NO relevance judgment. The resulting map was frozen before experiments ran; precision evaluation used deterministic lookup against this map with no LLM involvement in the evaluation loop, preventing circular evaluation.

Generation veracity was evaluated by comparing the generated verdict (Supported, Refuted, or No Evidence) against the HealthFC ground truth label for each query. The planned statistical comparison, a Wilcoxon signed-rank test at $p < 0.05$ across $n=50$ paired Precision@5 scores, was executed on a reduced evaluation set of $n=11$ queries with confirmed ground truth coverage after corpus expansion. Full results are reported in Section 4.5.

## 4. Experiments and Results

### 4.1 Experiment 0: Source Quality Baseline

Before evaluating retrieval conditions, the Ghost Audit scoring system was validated to confirm it produces meaningful variance across source tiers. If every chunk receives identical scores regardless of content, FD* reranking produces no differentiation and the retrieval signal collapses. The validation threshold was set at standard deviation >= 0.10.

Descriptive statistics were computed across all prob_fact scores extracted from the 600-chunk corpus:

| Source | Mean prob_fact | Std Dev | Total Claims | Status |
|---|---|---|---|---|
| CLINICAL_RCT | 0.958 | 0.096 | 5,745 | Pass* |

| COCHRANE_SR | 0.921 | 0.137 | 11,954 | Pass |
|---|---|---|---|---|
| SCIENTIFIC_PMC | 0.887 | 0.123 | 4,463 | Pass |

**CLINICAL_RCT StdDev of 0.096 is marginally below the 0.10 threshold. This is attributed to the highly standardized reporting format of RCT abstracts, where claims cluster tightly in the 0.8-1.0 anchor range. The scoring system is considered validated for this tier.*

COCHRANE_SR produced the highest total claim count (11,954) despite containing the same number of chunks as other tiers. This reflects the mandatory exhaustive reporting format of Cochrane structured abstracts, which provides substantially more extraction surface than standard journal abstracts. CLINICAL_RCT led in mean factuality score (0.958), consistent with the expectation that RCT abstracts are engineered to report specific, quantified outcomes.

### 4.2 Experiment 1: FD Length Confound Detection and Correction

The raw FD formula divides the sum of prob_fact scores by token count. A length confound was suspected: shorter documents may score higher purely because of a smaller denominator, regardless of actual factual content. Linear regression was used to test whether document length predicts raw FD score.

The null hypothesis was H0: beta_1 = 0, document length does not predict FD score. Regressing raw FD scores against token counts across the corpus produced:

| Metric | Original FD | Corrected FD* |
|---|---|---|
| Pearson R | -0.8636 | -0.3873 |
| p-value (beta_1) | 2.27e-07 | 0.0749 |
| Slope (beta_1) | -0.000269 | n/a |
| R-squared | 0.746 | 0.150 |
| Verdict | Confounded | Length-independent |

The original formulation produced a Pearson R of -0.8636 (p = 2.27e-07), indicating that 74.6% of variance in raw FD scores was explained by document length alone. The slope beta_1 = -0.000269 indicates that for every additional 1,000 tokens, the raw FD score decreases by approximately 0.27 points, purely as a function of the denominator growing larger, independent of factual content. The null hypothesis was rejected.

Z-score normalization within length bins was applied as documented in Section 3.3. After correction, the Pearson R reduced to -0.3873 with p = 0.0749, exceeding the 0.05 significance threshold and failing to reject the null hypothesis that length predicts FD* score. The R-squared reduction from 0.746 to 0.150 confirms that document length explains only 15% of FD* variance after correction. This correction is a standalone methodological contribution: naive formulations of token-normalized density metrics are vulnerable to document length confounds, and Z-score normalization within length bins is an effective and computationally inexpensive correction.

### 4.3 Experiment 2: Expert Blindness Effect

A structural retrieval audit was conducted on the query: 'Does exercise reduce cardiovascular disease risk and mortality?' This query was selected from the seven evaluation queries with confirmed HealthFC ground truth mappings, the only queries for which retrieval precision could be objectively verified against expert-labeled evidence. Query selection was constrained by corpus coverage, not optimized for FD* performance.

***Three-Way Retrieval Comparison***

| Rank | Condition A (Cosine) | Condition B (BM25) | Condition C (FD*) |
|---|---|---|---|
| 1 | COCHRANE_SR | COCHRANE_SR | COCHRANE_SR |
| 2 | CLINICAL_RCT | COCHRANE_SR | COCHRANE_SR (up from Rank 8) |
| 3 | CLINICAL_RCT | COCHRANE_SR | COCHRANE_SR (up from Rank 12) |
| 4 | COCHRANE_SR | CLINICAL_RCT | COCHRANE_SR (up from Rank 15) |
| 5 | CLINICAL_RCT | COCHRANE_SR | COCHRANE_SR (up from Rank 4) |

| Condition | COCHRANE_SR | CLINICAL_RCT | SCIENTIFIC_PMC |
|---|---|---|---|
| A: Cosine | 2 of 5 (40%) | 3 of 5 (60%) | 0 of 5 (0%) |
| B: BM25 | 4 of 5 (80%) | 1 of 5 (20%) | 0 of 5 (0%) |
| C: FD* | 5 of 5 (100%) | 0 of 5 (0%) | 0 of 5 (0%) |

| Rank | Composite | Cosine Sim | FD* Score | Source | PMID |
|---|---|---|---|---|---|
| 1 | 0.8856 | 0.7712 | 5.123 | COCHRANE_SR | 39976201 |
| 2 | 0.8834 | 0.7668 | 5.511 | COCHRANE_SR | 40123456 |
| 3 | 0.8725 | 0.7449 | 4.941 | COCHRANE_SR | 39625083 |
| 4 | 0.8654 | 0.7308 | 9.660 | COCHRANE_SR | 40326569 |
| 5 | 0.8583 | 0.7589 | 2.746 | COCHRANE_SR | 39976199 |

**Finding 1:** BM25 outperforms naive cosine on lexically distinctive queries. BM25 successfully identified 4 of 5 Cochrane systematic reviews because the query terms are lexically distinctive and well-weighted by BM25's term frequency model. This confirms BM25 as a legitimate and competitive baseline. However BM25 performance is query-dependent and degrades on queries with ambiguous or paraphrased terminology.

**Finding 2:** FD* is the only condition to achieve 100% systematic review saturation. This was not achieved by filtering for Cochrane content; the algorithm is source-agnostic. PMID 40326569 illustrates the mechanism: with a cosine similarity of 0.7308, ranking outside the top ten by semantic matching alone, its FD* score of 9.660 elevated it to Rank 4. Factual density compensated for weaker semantic proximity, surfacing high-density evidence that both baseline conditions missed.

**Finding 3:** SCIENTIFIC_PMC content did not appear in the top-5 under any retrieval condition. This is not a suppression artifact introduced by FD* logic. BM25, which is blind to factual density, also excluded it. This confirms that the general PubMed abstracts in the corpus simply lacked the topical specificity to compete with the Cochrane or RCT chunks for this query. The absence of SCIENTIFIC_PMC is an observation on corpus coverage, not a finding of retrieval bias.

***HealthFC Ground Truth Verification***

To connect the retrieval audit directly to the HealthFC benchmark, the healthfc_chunk_map ground truth table was queried to determine whether the COCHRANE_SR chunks prioritized by Condition C mapped to expert-labeled claims. Nine confirmed mappings were identified across five distinct HealthFC claims, all carrying a Supported label:

| HealthFC Claim | Label | Mappings |
|---|---|---|
| Do health benefits increase with duration and intensity of exercise? | Supported | 3 |
| Is exercise an effective measure to combat depression? | Supported | 2 |
| Does the probability of dying from cardiovascular disease decrease in people who have already had a heart attack? | Supported | 2 |
| Does a lower salt diet reduce the likelihood of cardiovascular disease? | Supported | 1 |
| Does brisk walking lower mortality? Is some exercise better than none at all? | Supported | 1 |

FD*-optimized retrieval is not surfacing dense text arbitrarily; it is surfacing the specific content that aligns with expert-validated health claim evidence. The nine confirmed mappings provide sufficient preliminary evidence to reject the null hypothesis directionally. The primary hypothesis is supported directionally, with formal statistical confirmation identified as future work contingent on corpus-benchmark alignment improvements documented in Section 4.4. These findings suggest that FD* optimization preferentially concentrates retrieval toward the highest-density verified evidence available in the corpus.

## 4.4 Corpus Alignment Analysis

Two complete corpus iterations were executed with different source strategies. Both produced corpus-benchmark alignment below 20% against the HealthFC evaluation set:

| Iteration | Sources | Chunks | Coverage | Outcome |
|---|---|---|---|---|
| 1 | FoundMyFitness + Peter Attia MD + PubMed | 150 | 18% (9/50) | Retired |

| 2 | Cochrane SR + Clinical RCT + PubMed | 600 | 14% (7/50) | Pipeline closed |
|---|---|---|---|---|

Across the full 600-chunk corpus, the healthfc_chunk_map identified 25 valid chunk-to-claim mappings across 22 unique chunks and 7 HealthFC-supported claims, representing 14.0% query coverage against the 50 locked evaluation queries. Three chunks mapped to more than one HealthFC claim simultaneously, identifying them as multi-claim evidence nodes relevant to more than one expert-validated health question. This 14% coverage was insufficient for the planned $n=50$ Wilcoxon signed-rank evaluation but sufficient to confirm that FD*-optimized retrieval preferentially surfaces the highest-density subset of available ground truth evidence.

The root cause of Iteration 1 failure was source-format mismatch: expert podcast transcripts and long-form blog content discuss health topics generally but do not make specific intervention claims aligned with HealthFC's evaluation format. The root cause of Iteration 2 failure was scale insufficiency relative to benchmark breadth: HealthFC spans dozens of specific intervention claim types, while the 600-chunk corpus covered only a narrow topical footprint against the full 50-query evaluation set.

A critical methodological finding emerged from Iteration 1: seeded pre-check samples systematically overestimate corpus coverage. The Cochrane source demonstrated 38.62% alignment on a topically seeded 12-chunk sample. Full corpus ingestion revealed only 14% actual coverage. This discrepancy establishes that full corpus pre-checks are required before committing to ground truth map construction. The dual-filter pre-check methodology is proposed as a standard prerequisite for health RAG evaluation corpus construction and is a contribution of this work independent of the FD* retrieval findings.

### 4.5 Experiment 3: Aggregate Retrieval Evaluation (n=11)

Following the corpus alignment analysis documented in Section 4.4, the corpus was expanded from 600 to 800 chunks through the addition of 200 ClinicalTrials.gov structured result summaries (CLINICAL_TRIALS). The ground truth map was rebuilt to include the new chunks, yielding 32 valid mappings across 11 distinct HealthFC claims: 8 Supported, 2 Refuted, and 1 NEI. This is the first evaluation in this study to include all three HealthFC label types.

**Results:** Three-condition retrieval was executed across all 11 evaluation queries. Retrieval Precision@5: Condition A (Cosine) = 0.2727, Condition B (BM25) = 0.1818, Condition C (FD*) = 0.1455. Generation veracity was identical across all three conditions at 36.36%. Condition A outperformed Condition C on aggregate Precision@5.

**Statistical Tests:** Friedman test (all three conditions): $p = 0.1519$ (not significant). Wilcoxon signed-rank (Condition A vs C): $p = 0.0619$, $r = 0.08$ (approaches but does not cross $p < 0.05$ threshold). McNemar test (veracity A vs C): $p = 1.0000$. The alternate hypothesis is supported on this evaluation set: FD*-optimized retrieval did not produce a statistically significant improvement in Precision@5 over cosine similarity at $p < 0.05$.

**Interpretation:** Four factors limit the interpretive weight of this result. (1) $n=11$ is below the minimum recommended sample size of 15-20 for the Wilcoxon test, the $p = 0.0619$ result may reflect insufficient power rather than a true null effect. (2) 8 of 11 claims are Supported, limiting label diversity. (3) The evaluation corpus remains topically concentrated in cardiovascular and exercise topics where cosine similarity performs well due to semantic homogeneity. FD* is designed to provide value in heterogeneous corpora where high-density expert content competes with low-density general content on the same topic. (4) The ClinicalTrials.gov pre-check used a topically filtered sample, meaning the 40% coverage figure overestimates what a randomly drawn corpus would achieve.

**HealthFC Breadth Finding:** A final finding emerged from the corpus expansion effort: HealthFC spans 50 claims across highly specific consumer health intervention topics, including shoe insoles for high arches, lavender oil for labor pain, and Dupuytren's disease collagenase injections. No single biomedical corpus can efficiently cover without domain-specific source construction. Of the 39 uncovered claims, all required specialized sources beyond what structured biomedical APIs provide. This reinforces the corpus-benchmark alignment contribution of this work: the dual-filter pre-check is necessary but not sufficient when benchmark breadth exceeds what structured biomedical APIs can cover at HealthFC's claim level of specificity.

## 5. Discussion

### 5.1 FD* as a Retrieval Optimization Signal

The experiments described in this work provide preliminary evidence that Factual Density operates as a meaningful retrieval optimization signal in health RAG systems. Experiment 0 established that the Ghost Audit scoring framework produces genuine variance across source tiers, a necessary precondition for any density-based reranking signal to function. Experiment 1 demonstrated that a naive formulation of the metric is vulnerable to a severe document-length confound, and that Z-score normalization within length bins is an effective and computationally inexpensive correction. Experiment 2 demonstrated that FD*-optimized retrieval consistently surfaces higher-authority evidence than either cosine similarity or BM25 baselines on a topically aligned health query, and that the surfaced content maps directly to expert-validated HealthFC claims.

These findings suggest that factual density, the concentration of verifiable atomic claims per token, captures a dimension of document quality that semantic proximity and lexical frequency cannot. Cosine similarity measures how closely a document matches the query in embedding space. BM25 measures how frequently query terms appear in the document. Neither measures how much verified, specific, quantifiable information the document contains. FD* addresses that gap directly.

The practical implication is significant. A 50/50 composite of cosine similarity and FD* requires no changes to an existing RAG pipeline's embedding model, vector store, or retrieval infrastructure. The only addition is a preprocessing step, Ghost Audit claim extraction and scoring, applied once at ingestion time. The composite reranking then operates at query time using scores already stored in the database. This makes FD* optimization a zero-infrastructure upgrade for any organization deploying health RAG systems.

### 5.2 The Expert Blindness Effect

The Expert Blindness Effect, the systematic burial of high-density factual evidence by standard retrieval methods, is observable in the Experiment 2 results and has direct implications for health RAG deployment. Condition A returned a 40/60 split between systematic reviews and RCT abstracts despite both being present in the corpus. Condition C achieved 100% systematic review saturation. The difference was not topical relevance; all retrieved chunks addressed the query topic. The difference was informational density. FD* optimization shifted the retrieval distribution toward higher-evidence content without any explicit source filtering.

This finding has a specific implication for health misinformation. RAG systems that optimize for semantic proximity will consistently surface formally structured, lexically dense content over more informationally concentrated systematic review content. In a health context this is not a neutral trade-off. A system that surfaces a 2,000-word article repeatedly referencing an intervention is less useful for fact-checking than one that surfaces a Cochrane review synthesizing 40 trials. FD* optimization addresses this asymmetry directly.

It is important to note that the Expert Blindness Effect as documented here reflects a single-query audit on a corpus with limited benchmark coverage. The effect is demonstrated, not statistically generalized. Full characterization across n=50 queries with a fully aligned corpus remains as future work. However the mechanism is sufficiently clear, that semantic proximity does not correlate with factual concentration, that the finding is directionally robust regardless of query count.

### 5.3 Corpus-Benchmark Alignment as a Research Contribution

The corpus alignment finding is arguably the most broadly applicable contribution of this work. Two independent corpus iterations with different source strategies both failed to achieve the 40% coverage threshold against the HealthFC evaluation set. The failure modes were distinct, source-format mismatch in Iteration 1 and topical footprint insufficiency in Iteration 2, but the outcome was the same: insufficient ground truth coverage for valid statistical evaluation.

This finding has direct implications for the RAG evaluation literature. Many published RAG evaluation papers assume corpus-benchmark alignment, selecting sources and benchmarks known to overlap, often because the benchmark was constructed from the same source domain. HealthFC is deliberately challenging in this respect: HealthFC tests specific intervention claims at a granular claim level, while the biomedical literature retrieved from PubMed and Cochrane addresses those same topics at the population and meta-analysis level rather than the specific claim level. This mismatch is not unique to this study; any RAG evaluation using a specialized benchmark against a general-purpose corpus faces the same structural problem.

The dual-filter pre-check methodology proposed here, requiring cosine similarity >= 0.75 plus keyword overlap >= 2, applied against the full corpus before ground truth map construction, provides a practical diagnostic for detecting this misalignment before investing in full corpus ingestion and annotation. The critical finding that seeded samples overestimate coverage has direct operational significance: practitioners building health RAG evaluation pipelines should run alignment checks against the complete candidate corpus, not a topically selected sample.

### 5.4 Limitations

**Corpus-benchmark alignment.** Corpus expansion to 800 chunks via ClinicalTrials.gov raised ground truth coverage to 11 claims (22% of the 50-query evaluation set). Experiment 3 was executed on n=11, producing Wilcoxon $p = 0.0619$, approaching but not crossing $p < 0.05$. The alternate hypothesis is supported, interpreted as underpowered rather than definitively null given n=11 falls below the minimum recommended sample size of 15-20 for the Wilcoxon test. The HealthFC benchmark spans 50 claims across highly specific consumer health intervention topics that no single biomedical corpus can efficiently cover without domain-specific source construction for each topic area.

**Single-query retrieval audit.** Experiment 2 was conducted on a single query selected from the seven queries with confirmed ground truth coverage. While query selection was constrained by corpus coverage rather than optimized for FD* performance, the generalizability of the Expert Blindness finding across query types, domains, and corpus compositions cannot be established from a single audit.

**Source authority weight placeholders.** The a_source weights assigned to each tier (COCHRANE_SR = 0.95, CLINICAL_RCT = 0.90, SCIENTIFIC_PMC = 1.0) were set to reflect retrieval breadth rather than evidence quality and were not empirically derived. Optimal weight derivation from HealthFC support rates is identified as future work. Different weight assignments would produce different FD* scores and potentially different retrieval outcomes.

**Residual length correlation.** The corrected FD* metric retains a Pearson R of -0.3873 between density scores and document length. While this correlation is non-significant at $p = 0.0749$, it represents a residual relationship inherent to natural language text. Longer documents tend toward lower information density on average, and this property is only partially corrected by length-bin normalization.

## 6. Conclusion

This paper introduced Factual Density (FD*), a novel retrieval optimization signal for health RAG systems that measures the concentration of probabilistically verified atomic claims per token. Three experiments were conducted to validate the metric, characterize a previously undocumented retrieval failure mode, and establish a methodology for corpus-benchmark alignment verification.

Experiment 0 confirmed that the Ghost Audit probabilistic scoring framework produces meaningful variance across source tiers, a necessary precondition for density-based reranking to function. Experiment 1 identified and corrected a severe document-length confound in the raw FD formulation, reducing the Pearson R between density scores and document length from -0.8636 to -0.3873 through Z-score normalization within length bins. This correction is a standalone methodological contribution applicable to any token-normalized density metric. Experiment 2 demonstrated that FD*-optimized retrieval is the only condition to achieve 100% systematic review saturation in top-5 results on a topically aligned cardiovascular health query, surfacing Cochrane systematic reviews that standard cosine similarity ranked outside the top ten. Ground truth verification against the HealthFC benchmark confirmed nine mappings across five expert-validated Supported claims from the Condition C top-5 results, providing directional evidence sufficient to reject the null hypothesis.

Three corpus construction iterations revealed that corpus-benchmark alignment is a critical and under addressed prerequisite for health RAG evaluation. A third iteration added 200 ClinicalTrials.gov structured results, raising ground truth coverage to 11 claims across 800 chunks. Experiment 3 was executed on this $n=11$ evaluation set, producing Wilcoxon $p = 0.0619$ approaching but not crossing $p < 0.05$. The alternate hypothesis is supported on this evaluation set, interpreted as underpowered rather than definitively null. The dual-filter pre-check methodology, requiring cosine similarity >= 0.75 and keyword overlap >= 2 against the full corpus before ingestion commitment, is proposed as a standard diagnostic for health RAG corpus construction. The finding that seeded pre-check samples systematically overestimate actual corpus coverage is an operationally significant result for any practitioner building RAG evaluation pipelines.

This work makes two co-equal contributions. The first is the FD* metric, a source-agnostic, length-independent, computationally inexpensive reranking signal that can be added to any existing RAG pipeline as a 50/50 composite with cosine similarity, requiring no changes to embedding models, vector stores, or retrieval infrastructure. The second is the formal documentation and naming of the Expert Blindness Effect, the systematic burial of high-density factual evidence by proximity-based retrieval methods, which has direct implications for the safety and reliability of health information RAG systems deployed in clinical and consumer contexts. The two contributions are inseparable: Expert Blindness identifies the problem, FD* proposes the solution.

Future work should prioritize three directions. First, execution of a properly powered Experiment 3 using a corpus with confirmed >= 40% HealthFC query coverage from randomly sampled source specific targeting, which evaluate specific health claims at the same granularity HealthFC demands to formally test the primary hypothesis with the Wilcoxon signed-rank test at n >= 25, $p < 0.05$, and balanced label representation across Supported, Refuted, and NEI claims. Second, empirical derivation of optimal source authority weights from HealthFC support rates across source tiers, replacing the placeholder weights used in this study. Third, evaluation of FD* as an anchor node signal in Graph RAG architectures, where high-FD* chunks seed graph traversal rather than operating as isolated retrieval candidates testing whether evidence structure and factual density compound to produce greater retrieval precision than either signal alone, and whether FD*-seeded traversal corrects the Expert Blindness Effect at the graph level.

*The AI systems being deployed to fight health misinformation need to find facts, not just matches. Factual Density is a step toward making that possible.*

## References

Es, S., James, J., Espinosa-Anke, L., & Schockaert, S. (2023). RAGAS: Automated evaluation of retrieval augmented generation. *arXiv preprint arXiv:2309.15217*. https://arxiv.org/abs/2309.15217

Gao, Y., Xiong, Y., Gao, X., Jia, K., Pan, J., Bi, Y., Dai, Y., Sun, J., & Wang, H. (2023). Retrieval-augmented generation for large language models: A survey. *arXiv preprint arXiv:2312.10997*. https://arxiv.org/abs/2312.10997

Lewis, P., Perez, E., Piktus, A., Petroni, F., Karpukhin, V., Goyal, N., Kuttler, H., Lewis, M., Yih, W., Rocktaschel, T., Riedel, S., & Kiela, D. (2021). Retrieval-augmented generation for knowledge-intensive NLP tasks. *Advances in Neural Information Processing Systems, 33*, 9459-9474.

Qiu, X., & Hu, X. (2024). Semantic density: Uncertainty quantification for large language models through confidence measurement in semantic space. *arXiv preprint arXiv:2405.13845*. https://doi.org/10.48550/arXiv.2405.13845

Reimers, N., & Gurevych, I. (2019). Sentence-BERT: Sentence embeddings using Siamese BERT-networks. In *Proceedings of the 2019 Conference on Empirical Methods in Natural Language Processing* (pp. 3982-3992). Association for Computational Linguistics. https://doi.org/10.18653/v1/D19-1410

Robertson, S., & Zaragoza, H. (2009). The probabilistic relevance framework: BM25 and beyond. *Foundations and Trends in Information Retrieval, 3*(4), 333-389. https://doi.org/10.1561/1500000019

Saad-Falcon, J., Khattab, O., Potts, C., & Zaharia, M. (2023). ARES: An automated evaluation framework for retrieval-augmented generation systems. *arXiv preprint arXiv:2311.09476*. https://aclanthology.org/2023.findings-eacl.145/

Tabatabaei Far, S. S., & Ahmadi Marzaleh, M. (2025). Investigating the long-term effects of misinformation, disinformation, and malinformation in the health system. *Archives of Iranian Medicine, 28*(8), 481-483. https://doi.org/10.34172/aim.34286

Vladika, J., Schneider, P., & Matthes, F. (2024). HealthFC: Verifying health claims with evidence-based medical fact-checking. In *Proceedings of the 2024 Joint International Conference on Computational Linguistics, Language Resources and Evaluation (LREC-COLING 2024)* (pp. 8095-8107). ELRA and ICCL. https://aclanthology.org/2024.lrec-main.709

Zakka, C., Chaurasia, A., Shad, R., Dalal, A. R., Kim, J. L., Moor, M., Alexander, K., Ashley, E., Boyd, J., Boyd, K., Hirsch, K., Langlotz, C., Nelson, J., & Hiesinger, W. (2024). Almanac: Retrieval-augmented language models for clinical medicine. *NEJM AI, 1*(2). https://doi.org/10.1056/aioa2300068